\documentclass[final,3p,times]{elsarticle}

\usepackage{multirow,setspace,times,amssymb,amsmath,graphicx,color,rotating,subfigure,url}
\usepackage{lineno}
\usepackage{natbib}

\usepackage{epsf}
\usepackage{rotating}

\usepackage{graphicx,epsfig}% Include figure files
\usepackage{dcolumn}% Align table columns on decimal point
\usepackage{bm}% bold math

\begin{document}

\title{Epidemic Dynamics On Information-Driven Adaptive Networks}

\author[inst1,inst2]{Xiu-Xiu Zhan} 
\author[inst1]{Chuang Liu} 
\ead{liuchuang@hznu.edu.cn}
\author[inst3]{Gui-Quan Sun}
\author[inst1,inst4]{Zi-Ke Zhang} 
\ead{zhangzike@gmail.com}

\address[inst1]{Alibaba Research Center for Complexity Sciences, Hangzhou Normal University, Hangzhou, 311121, P. R. China}
\address[inst2]{Department of Mathematics, North University of China, Taiyuan, Shan'xi 030051, P. R. China}
\address[inst3]{Complex Sciences Center, Shanxi University, Taiyuan 030006, P. R. China}
\address[inst4]{Alibaba Research Institute, Hangzhou, 311121, P. R. China}

\begin{abstract}
Research on the interplay between \emph{the dynamics on the network} and \emph{the dynamics of the network} has attracted much attention in recent years. In this work, we propose an information-driven adaptive model, where disease and disease information can evolve simultaneously. For the information-driven adaptive process, susceptible (infected) individuals who have abilities to recognize the disease would break the links of their infected (susceptible) neighbors to prevent the epidemic from further spreading. Simulation results and numerical analyses based on the pairwise approach indicate that the information-driven adaptive process can not only slow down the speed of epidemic spreading, but can also diminish the epidemic prevalence at the final state significantly. In addition, the disease spreading and information diffusion pattern on the lattice give a visual representation about how the disease is trapped into an isolated field with the information-driven adaptive process. Furthermore, we perform the local bifurcation analysis on four types of dynamical regions, including healthy, oscillatory, bistable and endemic, to understand the evolution of the observed dynamical behaviors. This work may shed some lights on understanding how information affects human activities on responding to epidemic spreading.
%The information-driven adaptive process can not only slow down the speed of epidemic spreading, but can also diminish the epidemic prevalence at the final state significantly
\end{abstract}

\begin{keyword} Epidemic spreading \sep Information diffusion \sep  Adaptive model
\end{keyword}


\maketitle

\section{\label{S1:Intro}Introduction}
The spreading dynamic is one of the core issues in network science \cite{Lloyd-May-2001-Science,Danon-Ford-House-IPID-2011}, where most of the related research processes focus on epidemic spreading and information diffusion in recent years. Much of the work to date focuses on the analysis of these two processes independently, such as the spread of single contagion \cite{Pastor-Satorras-Castellano-Mieghem-Vespignani-2014-Arxiv} or concurrent diseases \cite{Sanz-Xia-Meloni-Moreno-PRX-2014,Newman-PRL-2005}, and the diffusion of various kinds of information (e.g., news \cite{Chen-Chen-Gunnell-Yip-2013-PlosOne}, rumor \cite{Domenico-Lima-Mougel-Musolesi-SR-2013}, innovation \cite{Montanari-Saberi-2010-PNAS}.). However, the epidemic spreading process is closely coupled with the corresponding disease information diffusion (or saying awareness) in the real world. For example, the Google Flu Trend (GFT) can detect regional influenza outbreaks with the online search for the disease-related information \cite{Lazer-Kennedy-King-Vespignani-Science-2014,Carneiro-Mylonakis-CID-2009}. During the severe acute respiratory syndrome (SARS) outbreak in China in 2003, overwhelming number of disease reports have helped people win the ''people's war against SARS" finally  \cite{WorldHealthOrganization-2003,Tai-Sun-NMS-2007}. It is obvious that disease information diffusion plays an important role in control of the epidemic outbreak, but it is not easy to quantitatively measure the strength of its impact \cite{Funk-Gilad-Watkins-Jansen-PNAS-2009}.

Nowadays, some modeling approaches on the interaction between epidemic spreading and disease information diffusion have been proposed \cite{Funk-Gilad-Watkins-Jansen-PNAS-2009,Funk-Salathe-Jansen-JRSI-2010,Granell-Gomez-Arenas-PRL-2013}. The fundamental assumption is that, when a disease breaks out, people may get the disease information from their friends or media before the advent of the epidemic and take some preventive measures to keep away from being infected \cite{Funk-Salathe-Jansen-JRSI-2010,Wang-Tang-Yang-Do-Lai-Lee-SR-2014,Granell-Gomez-Arenas-PRE-2014}. By depicting preventive measures as the reduction of transmitting probability \cite{Funk-Gilad-Jansen-JTB-2010,Sahneh-Chowdhury-Scoglio-SR-2012} or particular states of individuals (immune or vaccination) \cite{Zhang-Wu-Tang-Lai-SR-2014}, previous models showed that the disease information diffusion indeed inhibits the epidemic spreading significantly (reduce the epidemic prevalence and enhance the epidemic threshold) \cite{Funk-Salathe-Jansen-JRSI-2010,Wu-Fu-Small-Xu-Chaos-2012}. Therefore, the emergence of mutual feedback between information diffusion and epidemic spreading \cite{Funk-Gilad-Watkins-Jansen-PNAS-2009} exhibits the intricate interplay between those two types of spreading dynamics. It should be noted that such complex interacted spreading dynamics are all performed on fixed networks (or population) in the aforementioned studies. However, individuals would sometimes cut off the connections with the infected ones when they are aware of the disease, leading to the continuous evolution of network structure. Consequently, how to characterize the mutual spreading process on the dynamical networks is a crucial issue we want to address in this work.

Generally, the network dynamic researches could be classified into two lines: (i) one is treated as the \emph{dynamics of the network}, which focuses on the time evolution of network topological properties \cite{Barabasi-Albert-1999-Science,Watts-Strogatz-1998-Nature,Holme-Saramaki-PR-2012}; (ii) the other is considered as the \emph{dynamics on the network}, which concerns the dynamical process of the nodes (or interactions) on networks, such as the epidemic spreading and information diffusion process \cite{Pastor-Satorras-Vespignani-2001-PRL,Liu-Zhang-2012-CNSNS}, the evolutionary game \cite{Gracia-Lazaro-Ferrer-Ruiz-PNAS-2012} and so forth. Currently, adaptive process \cite{Gross-Dlima-Blasius-2006-PRL} becomes the most important model to describe the mutual interaction between temporal network topology and node dynamics. In the adaptive process of epidemic spreading dynamics, the susceptible individuals are allowed to protect themselves by rewiring their links from the infected neighbors to some other susceptible ones \cite{Gross-Blasius-JRSI-2008,Gross-Sayama-2009,Tunc-Shaw-PRE-2014}. Many researches indicate that segregating infected (or susceptible) individuals with the adaptive process is an efficient strategy to reduce the fraction of susceptible-infected ($SI$) interactions, as well as hindering the outbreak of the whole epidemic spreading process \cite{Zhou-Xia-2014-PA,Volz-Meyers-2007-PRSB,Wieland-Aquino-Nunes-2012-EPL}. In addition, abundant temporal behaviors are presented for illustrating the spreading dynamics on the adaptive network, such as the coexistence of multiple stable equilibrium and the appearance of an oscillatory region, which are absent in the spreading dynamics on static networks \cite{Gross-Dlima-Blasius-2006-PRL,Guo-Trajanovski-Bovenkamp-Wang-Mieghem-PRE-2013}. Besides the edge rewiring strategy, the link cutting or temporarily deactivating is also a commonly used rule \cite{Tunc-Shkarayev-Shaw-JSP-2013,Valdez-Macri-Braunstein-PRE-2012}.

In this work, we aim to understand the complicated interplay between the epidemic spreading and disease information diffusion on dynamical networks. We propose an information-driven adaptive process to model the dynamical networks, where those who have been informed of the emergence of disease can break their neighbouring connections to avoid further epidemic spreading. And epidemic spreading and disease information diffusion are described by the SI and SIS model respectively, where the disease information generation of the infected individuals is considered to form a mutual feedback loop between these two types of spreading dynamics \cite{Funk-Gilad-Jansen-JTB-2010}. Therefore, the effect of information diffusion on epidemic spreading could be interpreted by two aspects: (i) reduce the epidemic spreading probability with protective measures; and (ii) cut off $SI$ links with the information-driven adaptive process. Both numerical analyses based on the pairwise approach and simulation results indicate that the information diffusion and the adaptive process can inhibit the epidemic outbreak significantly. In addition, we present a full local bifurcation diagram to show the complex dynamical behaviors in the proposed model.

\section{Results}
\makeatletter
\newcommand{\rmnum}[1]{\romannumeral #1}
\newcommand{\Rmnum}[1]{\expandafter\@slowromancap\romannumeral #1@}
\makeatother

\begin{figure}[htb]
\centering
\includegraphics[width=10cm]{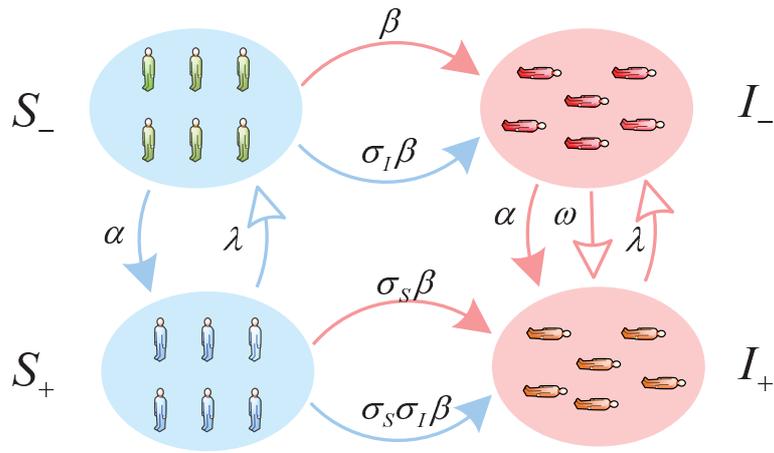}
\caption{\label{Fig1} (Color online) Transmission diagram of SI epidemic model and the information about the epidemic among the population.}
\end{figure}

\noindent{\textbf{Model description.}} In Fig. \ref{Fig1}, we illustrate the proposed model in detail. The vertical transformation describes the diffusion of disease information by an SIS model, where individuals can be at one of the two states: (i) $+$: indicates that the individuals have known the existence of the disease, denoted as the informed ones; (ii) $-$: indicates that the individuals have not known the existence of the disease. At each time step, the informed nodes will transmit the information to their unknown ($-$) neighbours with probability $\alpha$, and each informed individual can forget the information of the disease with a probability $\lambda$. Besides, the one who has been infected by the disease will become to know the information of the disease with a corresponding rate $\omega$ \cite{Funk-Gilad-Watkins-Jansen-PNAS-2009,Granell-Gomez-Arenas-PRL-2013}.

In the horizontal transformation of Fig. \ref{Fig1}, the epidemic spreading is described by an SI model. Each node is at one of two states, susceptible (S) or infected (I). The disease can be transmitted through the SI links, where the S-state individuals could be infected with the probabilities $\beta$, $\sigma_I\beta$, $\sigma_S\beta$ and $\sigma_{SI}\beta$ respectively through $S_-I_-$, $S_-I_+$, $S_+I_-$ and $S_+I_+$ links, where $\sigma_I, \sigma_S$ and $\sigma_{SI}$ are the impact factors of the information on  epidemic spreading. Generally, when people know the occurrence of the disease (informed individuals), they would like to take some measures to protect themselves, leading to the reduction in infectivity ($0<\sigma_S,\sigma_I<1$). In particular, the influence coefficient of the epidemic spreading probability through $S_+I_+$ links could be calculated as $\sigma_{SI}=\sigma_S\sigma_I$, with the assumption of the independent effect of the infected probability.

Additionally, we consider an information-driven adaptive process which the informed individuals would reduce physical contacts to protect themselves or their friends. That is to say, the informed susceptible individuals ($S_{+}$) will keep away from their infected neighbors to protect themselves from being infected, and informed infected individuals ($I_{+}$) will also avoid contacting their susceptible neighbors to prevent the epidemic from further spreading. Consequently, the edge-breaking rule of adaptation is adopted \cite{Tunc-Shkarayev-Shaw-JSP-2013}. Thus, at each time  step, the $S_+$ ($I_+$) state individuals will break the links connected to their $I$ ($S$)-state neighbors with probability $r_{S}$ ($r_I$) respectively. Specially, the breaking probability of the $S_+I_+$ pairs could be interpreted as $1-(1-r_S)(1-r_I)$ with the independent assumption. It should be worth noting that the deactivation of $SI$ links only represents the avoidance of physical contacts between the $S$- and $I$-state individuals. That is to say, the edge-breaking process will not affect the diffusion of disease information for it can be transmitted through other types of connections such as phone, internet and so forth. It is worth noting that the dynamic of the epidemic spreading is simply a classical SI model when the parameters are set as $r_S=r_I=0$, i.e., there is no edge-breaking in this case.

According to the model described above, the spreading process can be summarized as follows. At the beginning, an individual is randomly selected as the $I_{+}$ node, which is considered as the \emph{seed} of both the epidemic spreading and information diffusion, and all other individuals are set as $S_{-}$ ones. At each time step, (i) the infected individuals would transmit the disease to their susceptible neighbors with the corresponding probabilities; (ii) the informed individuals would transmit the disease information to their un-informed neighbors; (iii) the informed individuals can forget the information; (iv) the informed individuals would also break the links with their relevant neighbors by considering the adaptive mechanism. Finally, the spreading process would be terminated when the size of the infected individuals becomes stable.\\
% We perform the proposed model on a random network with network size $N=10000$, and average degree $\langle k\rangle=6$

\noindent{\textbf{Numerical mathematical analysis.}} Firstly, we develop theoritical analyses to depict the dynamic processes of both information diffusion and epidemic spreading. In particular, mean-field analysis and the pairwise analysis are adopted. Let $\chi$ be the state variable, thus $[\chi]$ denotes the expected values of individuals of different type at the population (e.g. $[S_{+}]$ and $[S_{+}I_{+}]$ represent the expected number of informed susceptible nodes and expected number of links connecting an informed susceptible node to an informed infected node respectively).

Therefore, with the classical mean-field approach, we can obtain:
\begin{equation}
\label{Eq:Meanfield:I+}
\frac{d[I_{+}]}{dt}=\langle k\rangle[S_{+}](\sigma_{S}\beta [I_{-}]+\sigma_{S}\sigma_{I}\beta [I_{+}]) +\alpha[I_{-}]([S_{+}]+[I_{+}])+\omega [I_{-}]-\lambda [I_{+}]
\end{equation}
comparatively, with the pairwise approach, we can obtain:
\begin{equation}
\label{Eq:Pairwise:I+}
\frac{d[I_{+}]}{dt}=(\sigma_{S}\beta[S_{+}I_{-}]+\sigma_{S}\sigma_{I}\beta[S_{+}I_{+}])+\alpha([S_{+}I_{-}]+[I_{-}I_{+}])+\omega[I_{-}]-\lambda[I_{+}]
\end{equation}
where, the first terms of Eq.(\ref{Eq:Meanfield:I+}) and (\ref{Eq:Pairwise:I+}) describe the infection of the $S_+$-state individuals, the second terms describe the information acceptance of the $I_-$-state individuals, the third terms describe the information generation of the $I_-$-state individuals and the last terms represent the information loss of the $I_+$-state individuals. Simultaneously, the full set of differential equations based on those two approaches can be illustrated in \textbf{Appendix}. By the way, the adaptive process could be described by the last terms of $\dfrac{d[S_+I_-]}{dt}$, $\dfrac{d[S_-I_+]}{dt}$ and $\dfrac{d[S_+I_+]}{dt}$ in the pairwise approach of Eq. (\ref{Eq:Pairwise:All}). It should be noted that the pairwise analysis is based on a well-known closure approximation given by $[ABC]=\dfrac{[AB][BC]}{[B]}$ with the assumption that the degree of each individual obeys Poisson distribution \cite{Morris-1997,Keeling-PRSB-1999}. In general, it might be very hard to get exact solutions of such complex differential equations, thus we give numerical solutions of the equations instead of the theoretical analysis in the following analysis.\\

\begin{figure}[htb]
\centering
\includegraphics[width=10cm]{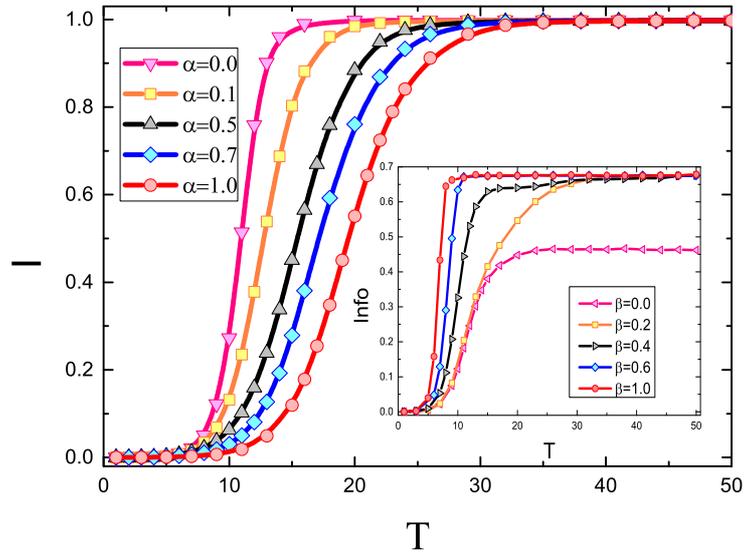}
\caption{\label{Fig2} (Color online) The epidemic spreading dynamics of various information diffusion probabilities $\alpha$ without considering the effect of adaptive process. The parameters are set as $\beta=0.3, \sigma_{S}=0.5, \sigma_{I}=0.7, \lambda=0.2, \omega=0.75, r_{S}=r_{I}=0.$ The inset shows the information diffusion dynamics ($Info$) of various $\beta$ for $\alpha=0.6$. }
\end{figure}

\noindent{\textbf{Simulation and numerical analysis without adaptive behaviour}}. We first consider a simple case of no adaptive process, i.e., there is no edge-breaking. Fig. \ref{Fig2} gives the simulation result of the epidemic spreading dynamics for various information diffusion probabilities $\alpha$, with the epidemic spreading probability $\beta=0.3$. For the SI process, the whole population would be infected when $\beta>0$ for the connected social networks, resulting in that the infected density equals to 1 at last for all the value of $\alpha$ in Fig. \ref{Fig2}. That is to say, the diffusion of the disease information couldn't reduce the size of the epidemic outbreak with the SI assumption. Whereas it is obvious to find that the epidemic spreading speed decreases with the increase of $\alpha$, where the time cost for the whole population becomes infected when $\alpha=1$ is about three times longer than that of $\alpha=0$. In this sense, the diffusion of the disease information can inhibit the epidemic spreading significantly. In addition, the inset of Fig. \ref{Fig2} indicates that the epidemic spreading can enhance the disease information diffusion. Actually, according to the transformation process in Fig. \ref{Fig1}, on the one hand, we can realize that the epidemic spreading could be influenced by information diffusion where the epidemic spreading probability of the informed individuals would change; and on the other hand, the information diffusion could be influenced by the epidemic spread where the social disease information level (namely $Info$ in the inset of Fig. \ref{Fig2}) would be higher if more people are infected for the information generation, denoted by the parameter $\omega$. In this way, a mutual feedback between disease spreading and information diffusion emerges: higher prevalence of the infected individuals keeps more disease information generation in the population, which in turn gives rise to more informed individuals, thereby weakening the spread of epidemic.

\begin{figure}[htb]
\centering
\includegraphics[width=10cm]{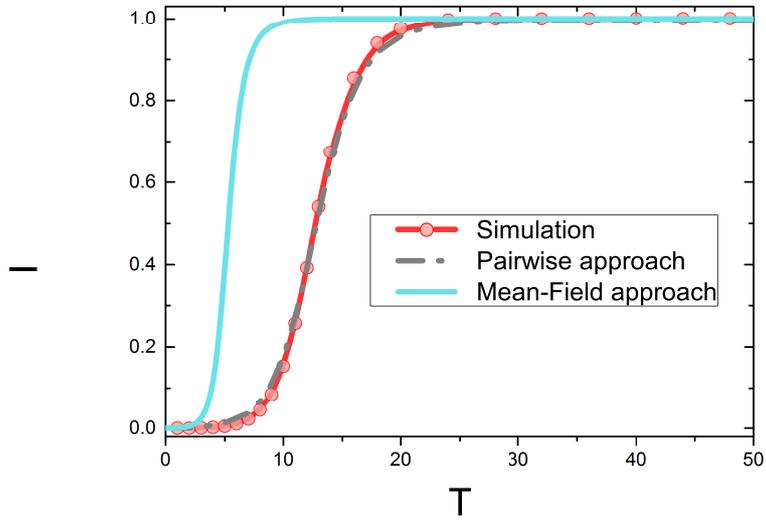}
\caption{\label{Fig3}  (Color online) Comparison of simulation results with the mean-field model and the pair approximation model without considering the effect of adaptive process. The parameters are set as $\beta=0.3, \sigma_{S}=0.5, \sigma_{I}=0.7, \lambda=0.2, \omega=0.75, \alpha=0.6, r_{S}=r_{I}=0$. }
\end{figure}

Fig. \ref{Fig3} presents the numerical analysis and the simulation result of the epidemic spreading dynamics when $r_S=r_I=0$, and the curves of the mean-field approach and pairwise approach are calculated according to Eq. (\ref{Eq:Meanfield:All}) and (\ref{Eq:Pairwise:All}) respectively. The epidemic spreading based on the classical mean-field approach is much quicker than that of the simulation result, which would be caused by the mean-field assumption on the SI model. In the mean-field assumption, the $I$- and $S$-state individuals are well-distributed in the system. However, in the SI process, the $I$-state individuals are all well clustered, resulting in that many $I$-state individuals have no chance to contact the $S$-state individuals. In this way, the classical mean-field approach can not exactly describe the SI dynamic process. However, such problem is not so significant in the pairwise approach, which consider the time evolution of the links directly. And Fig. \ref{Fig3} shows that the time evolution of the spreading process on the pairwise approach finds good agreement with the simulation result.\\

\begin{figure}[htb]
\centering
\includegraphics[width=8cm]{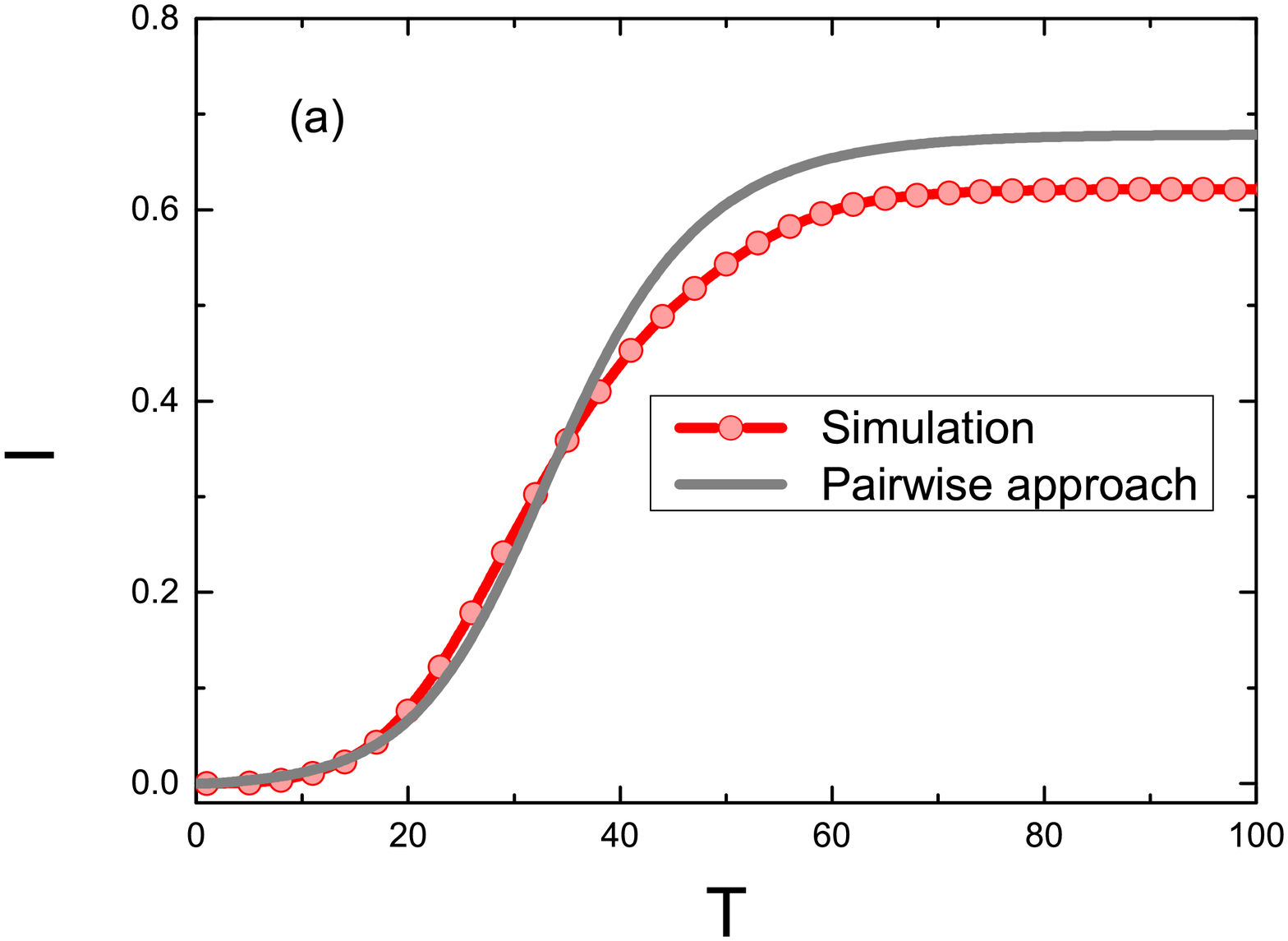}
\includegraphics[width=8cm]{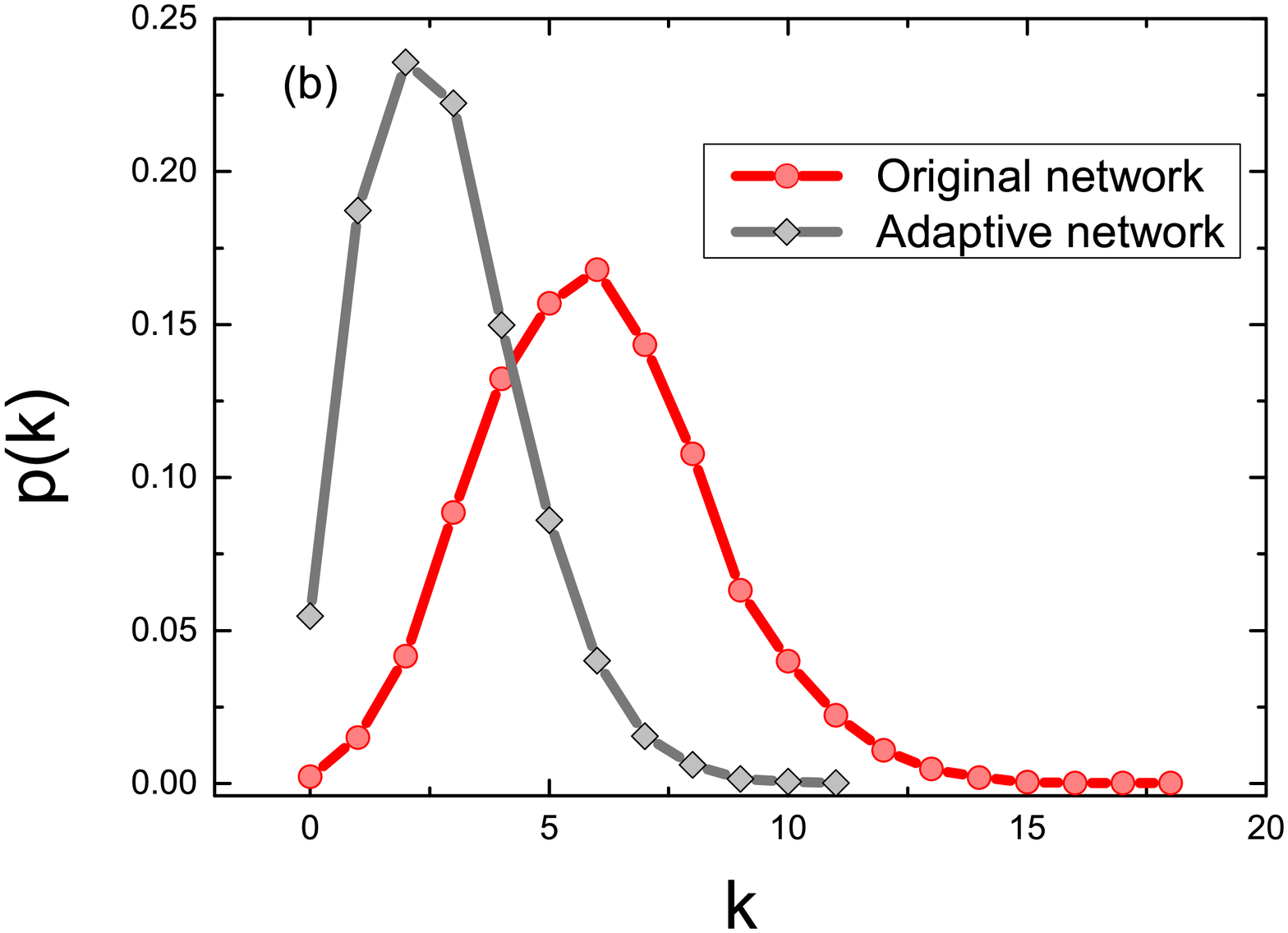}
\caption{\label{Fig4}  (Color online) Dynamical analysis of the spreading model with adaptive process. (a) Comparison of the pairwise model with the simulation results. (b) Degree distribution of the original network and that after the adaptive process. The parameters are set as $\beta=0.2, \sigma_{S}=0.5, \sigma_{I}=0.7, \lambda=0.2, \omega=0.2, \alpha=0.5, r_{S}=0.15, r_{I}=0.1$.}
\end{figure}

\noindent{\textbf{Spreading dynamics with the adaptive process}}. In this part, we shall present the spreading dynamics with the information-driven adaptive process, shown in Fig. \ref{Fig4}. Different from the results of Fig. \ref{Fig2}, the saturation value of the infected density at the final state is much smaller than 1 in Fig. \ref{Fig4}a. That is to say, with the adaptive process based on the information diffusion, many individuals could avoid being infected via reducing some contacts. In addition, we also plot the numerical solution based on the pairwise approach in Fig. \ref{Fig4}a. It can be seen that the pairwise solution is not so well for the spreading dynamic on the adaptive system. The difference might be caused by the network structure variation in the adaptive process, where the assumption of the pairwise approach is the Possion degree distribution. And this conjecture is proved in Fig. \ref{Fig4}b, where the degree distribution of the original network is approximate to the Possion-distribution (pink circle markers), while the distribution of the network at the final state (gray diamond markers) is quite different. In addition, Fig. \ref{Fig4}a shows that the difference becomes larger with the increase of time, where the degree distribution deviates more away from the Possion distribution when the process goes on.

\begin{figure}[htb]
\centering
\includegraphics[width=14cm]{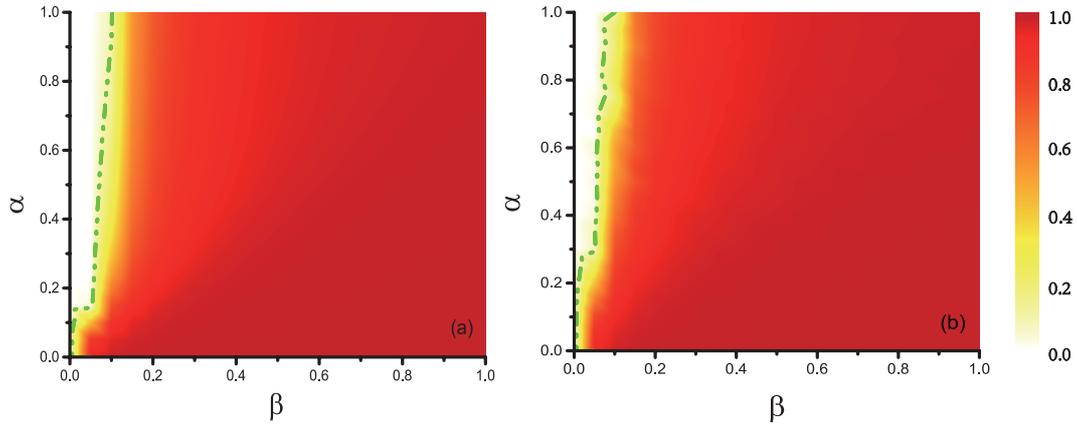}\\

\caption{\label{Fig5}  (Color online) The fraction of infected individuals in the stationary state (colors in the phase diagram represent the density of infected individuals at the final state, the dashed green curve shows that the prevalence value transmits from near 0 to significantly larger than 0) versus $\alpha$ and $\beta$ for (a) pairwise analysis and (b) simulation result. The parameters are set as  $\sigma_{S}=0.5, \sigma_{I}=0.7, \lambda=0.2, \omega=0.2, r_{S}=r_{I}=r=0.1$. }
\end{figure}

The information-driven adaptive process can not only slow down the speed of epidemic spreading, but can also diminish the epidemic prevalence at the final state significantly according to Fig. \ref{Fig2} and Fig. \ref{Fig4}. For simplicity, we assume $r_S=r_I=r$ in the following analysis. In order to exhibit the influence of information diffusion in detail, we show the full phase diagram $\alpha-\beta$ for $r=0.1$ in Fig. \ref{Fig5}. The Fig. \ref{Fig5}a and \ref{Fig5}b are the numerical solution of the pairwise approach and the simulation result, respectively. As stated previously, the numerical solution is not very precise, but it can match the overall trend of simulation result well. For a fixed epidemic spreading probability $\beta$, the size of the epidemic outbreak reduces with the increase of $\alpha$. That is to say, the disease information diffusion can inhibit the epidemic spreading. Analogously, the quicker and broader of the information diffusion (larger $\alpha$) is, the more efficient inhibition on the epidemic spreading will be. In addition, the curve of the color mutation (the dashed green curve) in Fig. \ref{Fig5} could be considered as the transition point, where the epidemic can't spread out if $\alpha$ and $\beta$ locate at the area on the left of this curve (the white range). The threshold value of the epidemic spread probability becomes larger with the increase of $\alpha$.

\begin{figure}[htb]
\includegraphics[width=4cm]{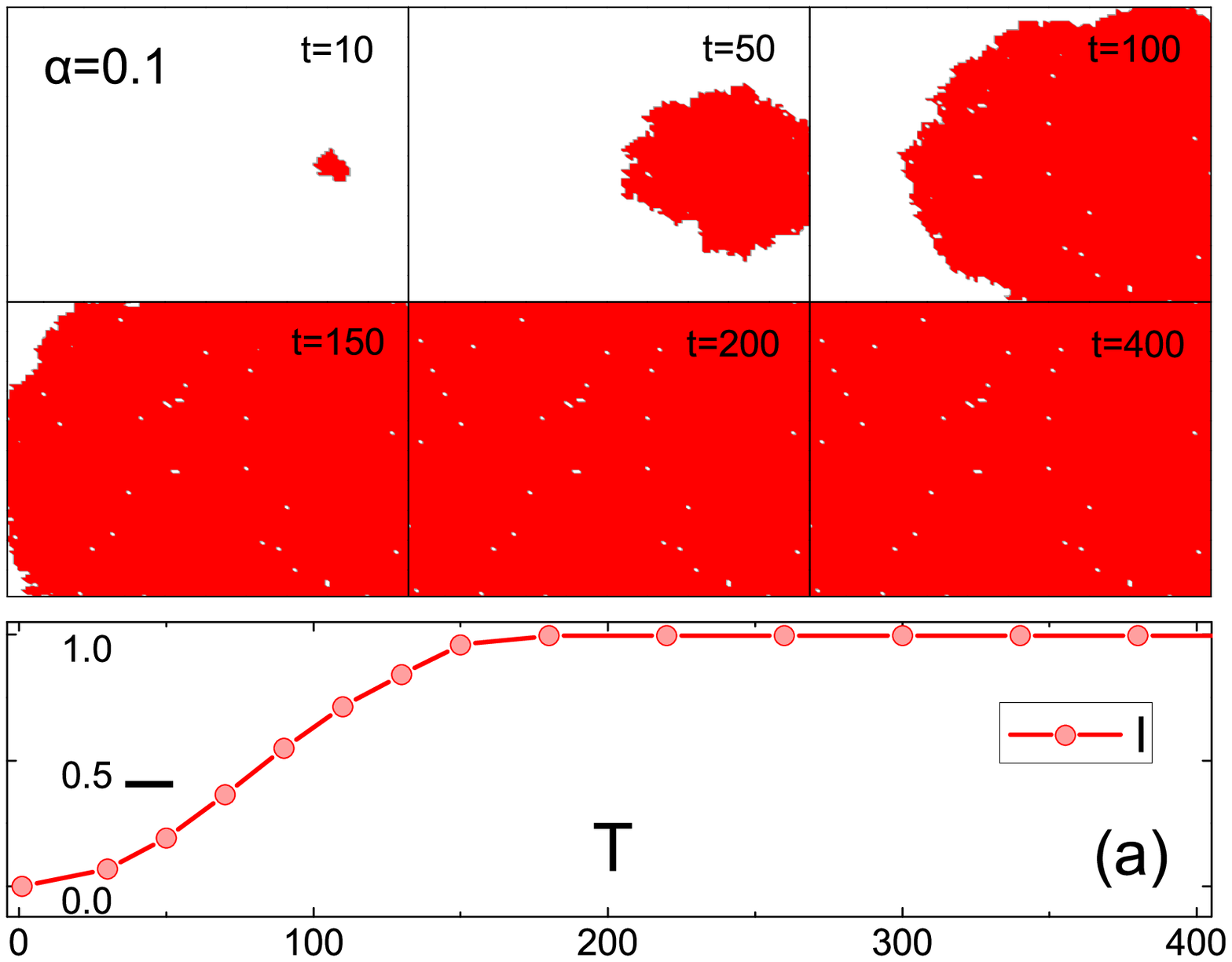}
\includegraphics[width=4cm]{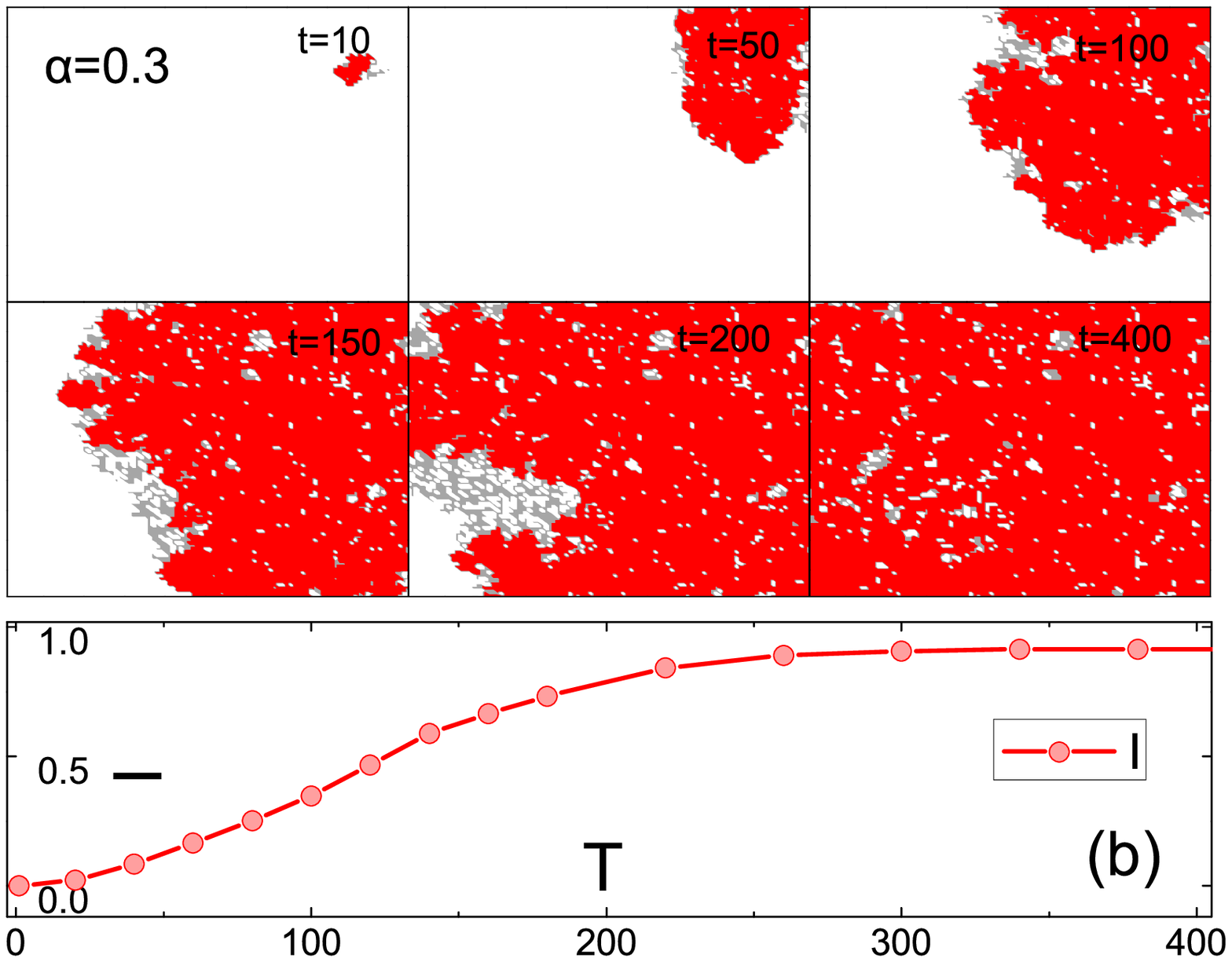}
\includegraphics[width=4cm]{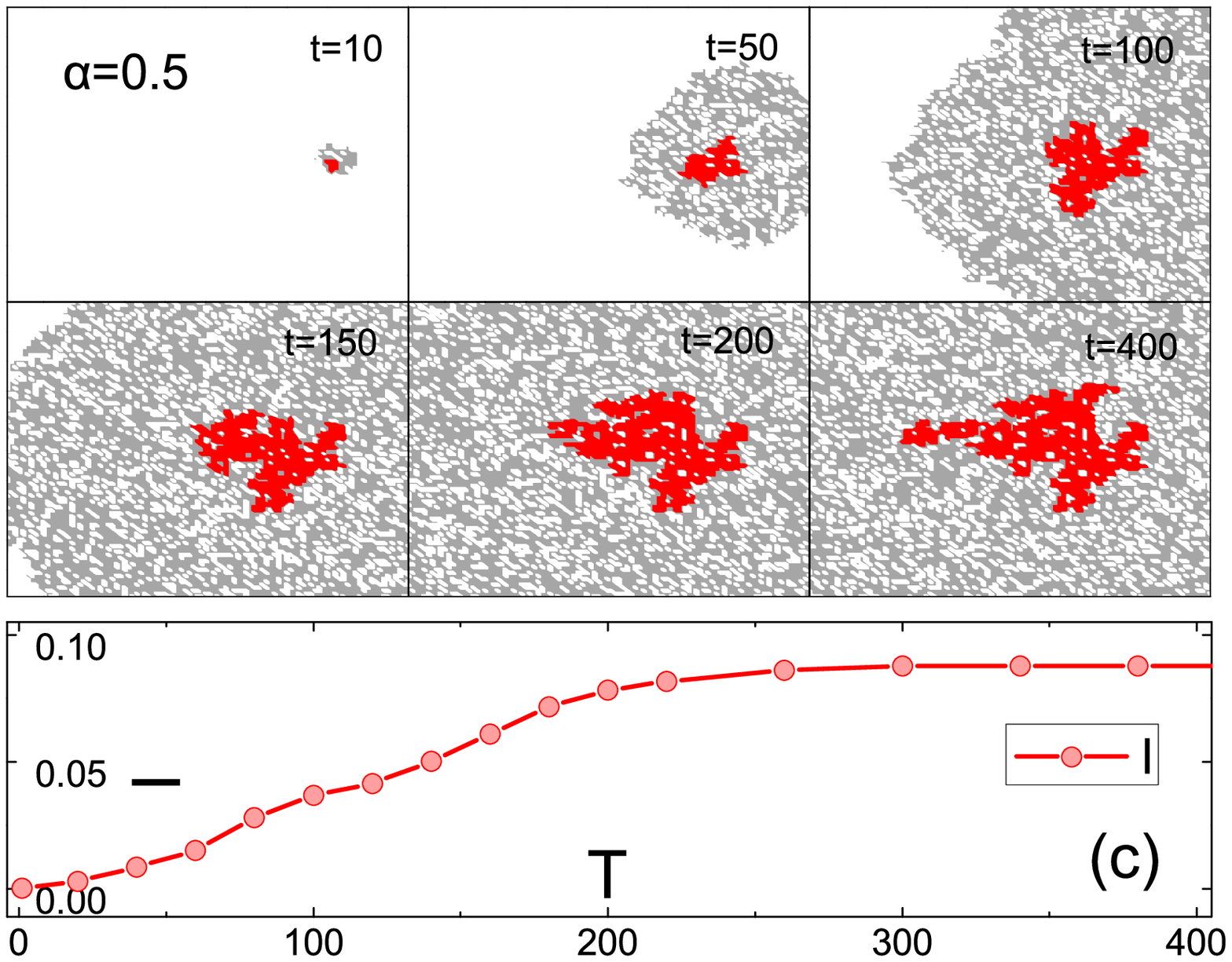}
\includegraphics[width=4cm]{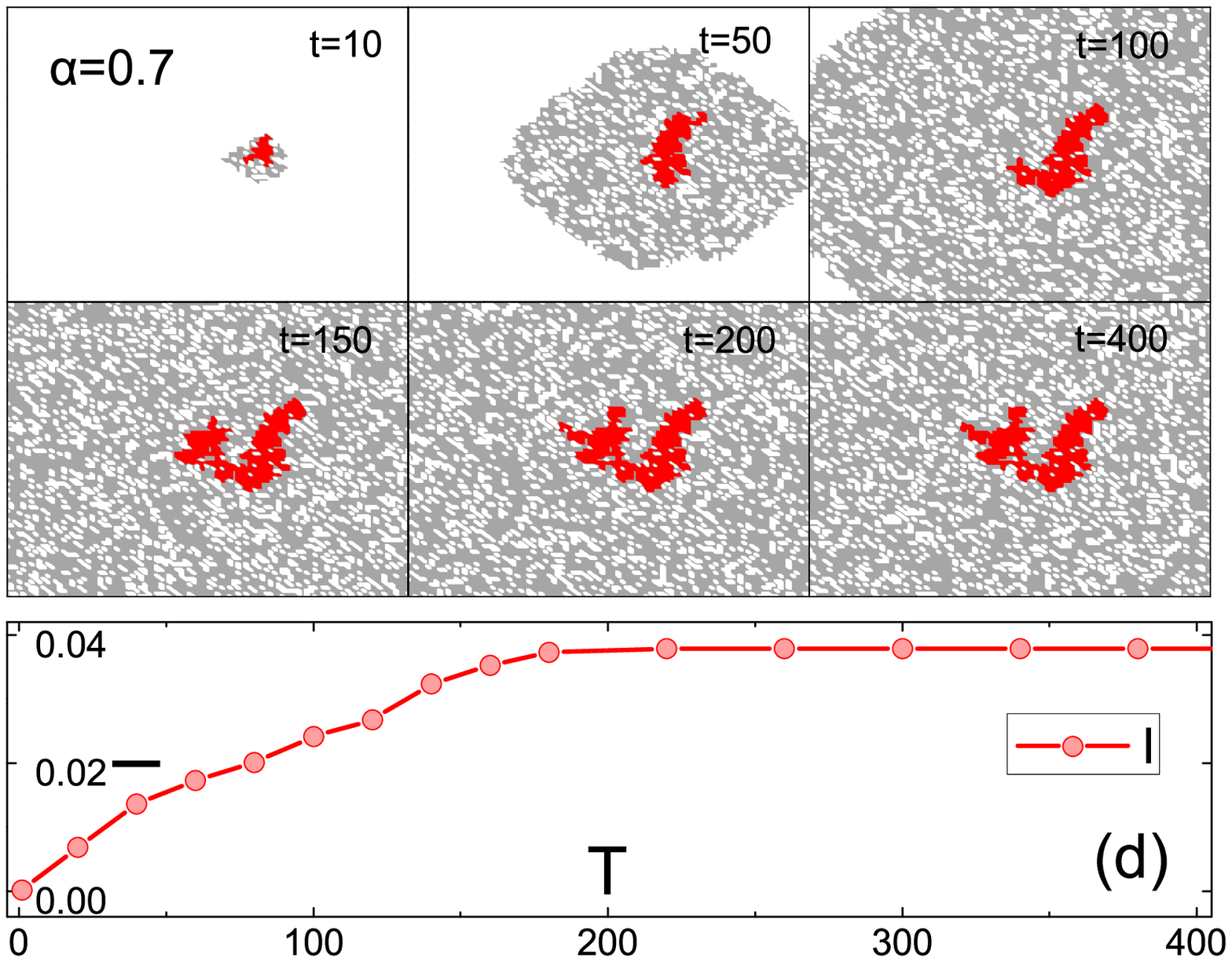}
\caption{\label{Fig6} (Color online) Illustration to dynamic spreading process by considering the adaptive effect on the lattice. The square gridding patterns show the distribution of the infected and informed individuals in some particular time steps. The red area represents the nodes that are infected by the epidemic, while the gray area represents the informed individuals. The red curves (lower panels) describe the fraction of infected individuals over time with corresponding. (a) $\alpha=0.1$; (b) $\alpha=0.3$; (c) $\alpha=0.5$; (d) $\alpha=0.7$. Other parameters are set as  $\beta=0.4, \sigma_{S}=0.4, \sigma_{I}=0.8, \lambda=0.1, \omega=0.2, r_{S}=r_{I}=r=0.1.$}
\end{figure}

In order to intuitively demonstrate the epidemic spreading and the information diffusion process, we show the simulation results of those two types of spreading processes on a $100\times100$ lattice in Fig. \ref{Fig6} for various $\alpha$. It presents four kinds of different levels of information spreading processes (corresponding to different $\alpha$), and observe how the information diffusion affects the spreading of epidemic. In addition, as the adaptive edge-breaking process is merely executed on the epidemic spreading process, while these edges can still transmit information, the prevalence of information diffusion could maintain at a high level. For each $\alpha$ in Fig. \ref{Fig6}, firstly we give the fraction of the infected individuals in each time step (the red curve in each subfigure). And for some particular time steps, we show the states of each individual with the gridding patterns, where the red dots and the gray dots represent the infected and informed individuals respectively (the contact networks and the un-informed susceptible individuals are not shown in the figures). We can intuitively see the distribution of the infected and informed individuals and conclude that when the diffusion of information is slower than the epidemic, we cannot stop the epidemic from spreading (Fig. \ref{Fig6}a and \ref{Fig6}b), however, when the information is diffusing faster, the epidemic will be trapped into an isolated area and cannot spread anymore (Fig. \ref{Fig6}c and \ref{Fig6}d).\\

\begin{figure}[htb]
\centering
\includegraphics[width=10cm]{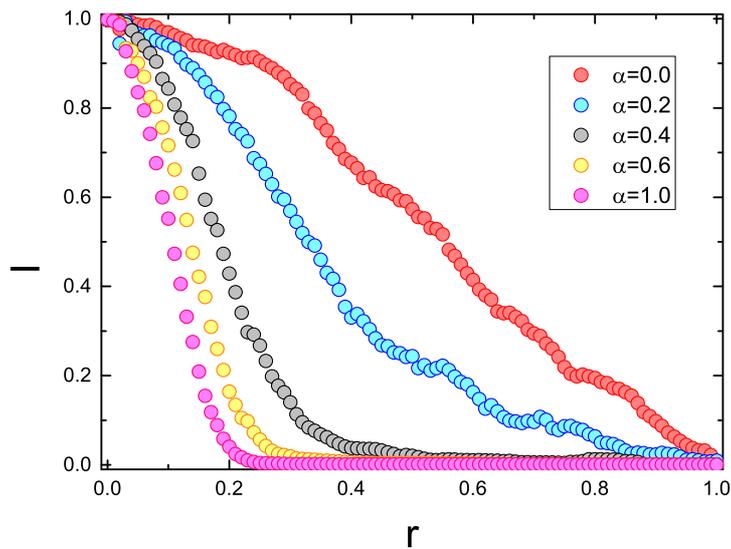}
\caption{\label{Fig7}  (Color online) Fraction of infected individuals versus $r$. Different curves correspond to different $\alpha$. Other parameters are set as $\beta=0.2, \sigma_{S}=0.5, \sigma_{I}=0.7, \lambda=0.2, \omega=0.2.$}
\end{figure}

\noindent{\textbf{The sensitivity of the edge-breaking probability on epidemic spreading dynamics}}. The phase diagram in Fig. \ref{Fig5} shows the impact of information diffusion rate $\alpha$ on the epidemic spreading dynamics. In general, the adaptive edge-breaking probability $r_S$ and $r_I$ are also important parameters in affecting the epidemic spreading process. Fig. \ref{Fig7} illustrates the epidemic prevalence in the final state versus the adaptive edge-breaking rate ($r$) for various information diffusion rate $\alpha$. It can be found that the epidemic prevalence diminishes with the increase of $r$, that is to say, the epidemic could be controlled if people are very sensitive with the disease information and subsequently keep away from the infected. It should be noted that there is no disease information diffusion when $\alpha=0$, but with considering the information generation, the infected individuals could stop contacting with the susceptible neighbors to impede the further spreading of epidemic. With the increase of $\alpha$, the epidemic prevalence reduces sharply versus $r$ and the continuous transition could be observed. By the way, it will change to a total isolation of infected individuals for $r=1$, which seems to be the most effective way in controlling the contagion \cite{Crokidakis-Queiros-JSM-2012,Lagorio-Dickison-Vazquez-Braunstein-PRE-2011}.\\

\begin{figure}[htb]
\centering
\includegraphics[width=13cm]{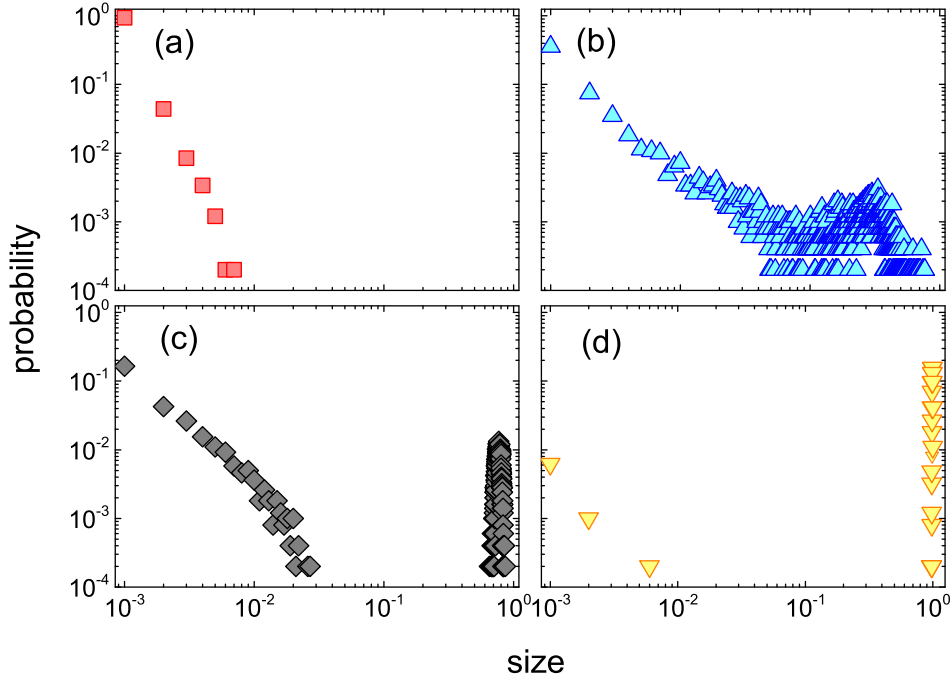}
\caption{\label{Fig8} (Color online) Distribution of the infected density in the final state versus different values of r and $\beta$.  Each distribution is obtained by carrying out 10,000 independent realizations for the final fraction of infected. The parameters are set as $r_{S}=r_{I}=r=0.7, 0.35, 0.15, 0; \beta=0.05, 0.35, 0.25, 0.4$ for (a), (b), (c) and (d) respectively. Other parameters are  $\sigma_{S}=0.5, \sigma_{I}=0.7, \lambda=0.2, \omega=0.2, \alpha=0.6$.}
\end{figure}

\noindent{\textbf{Abundant dynamical features based on information-driven adaptive process.}} In order to deeply characterize the complex dynamical features of the proposed process, we concentrate on the distribution of the infected density in the final state ($I^*$) rather than the simple average value \cite{Gross-Dlima-Blasius-2006-PRL,Guo-Trajanovski-Bovenkamp-Wang-Mieghem-PRE-2013}. Fig. \ref{Fig8} shows four different types of dynamical behavior by calculating the distribution of the final fraction of infected for various $\beta$ and $r$ . For the distribution of Fig. \ref{Fig8}a, we have carried out 10000 realizations of the infected density, and above $94\%$ of the infected density is 0.0001, and the maximal is 0.0007, i.e., the infected density  $I^*\rightarrow0$, thus we consider this distribution indicates a healthy state (the disease can't spread out) under the parameters setting here. Similarly, as to the distribution of Fig. \ref{Fig8}d, above $90\%$ of the infected density is higher than 0.8, indicates a case of endemic state (epidemic outbreaks). Whereas the case illustrated in Fig. \ref{Fig8}c is very different, where the infected density $I^*$ is around either zero or a nonzero value. This indicates that a bistable state \cite{Gross-Dlima-Blasius-2006-PRL} is located in this model, where healthy state and endemic state are both stable in this case. In addition, the oscillatory dynamic can also be observed in particular parameter settings (Fig. \ref{Fig8}b).

\begin{figure}[htb]
\centering
\includegraphics[width=5.7cm]{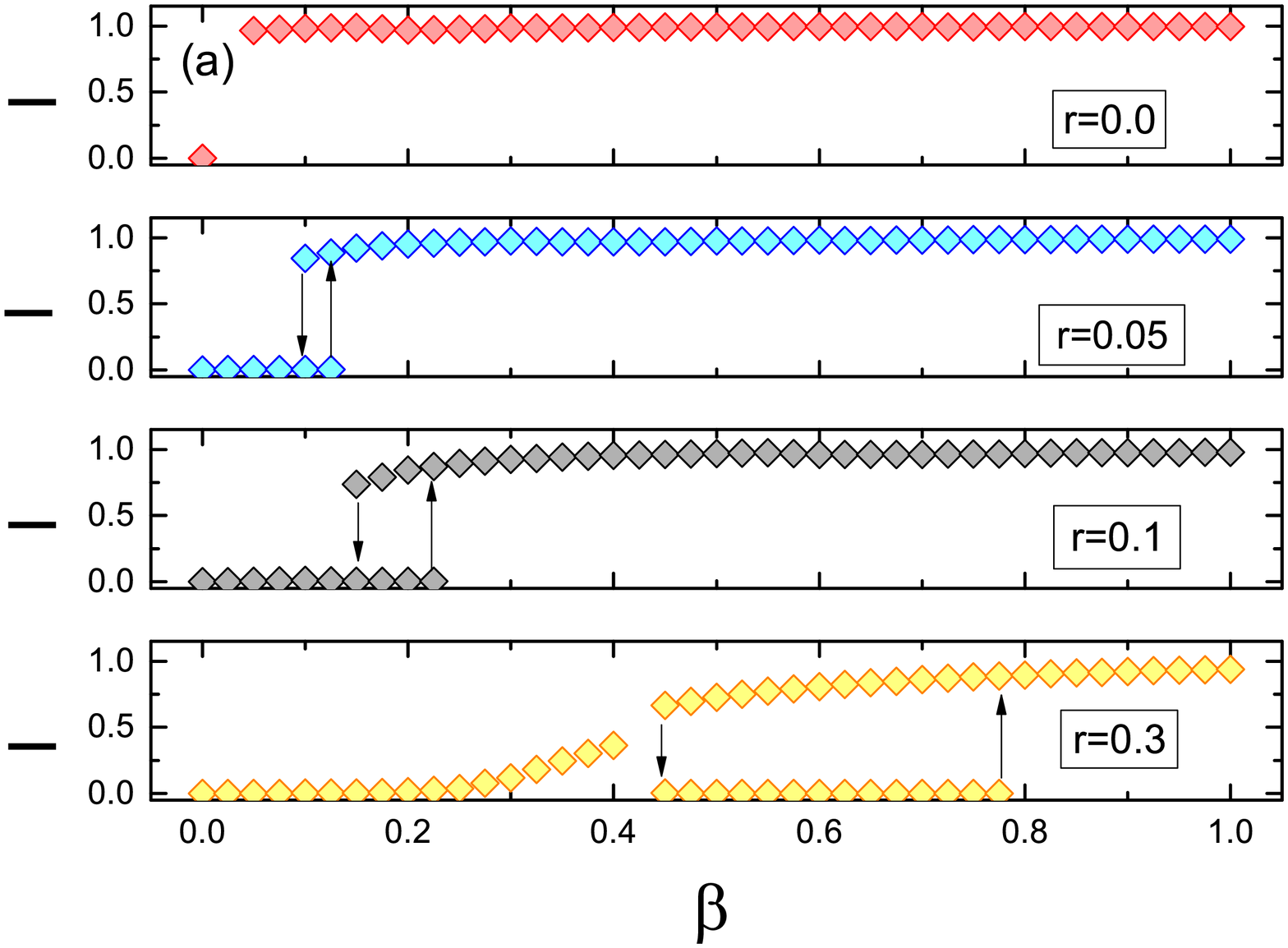}
\includegraphics[width=5.4cm]{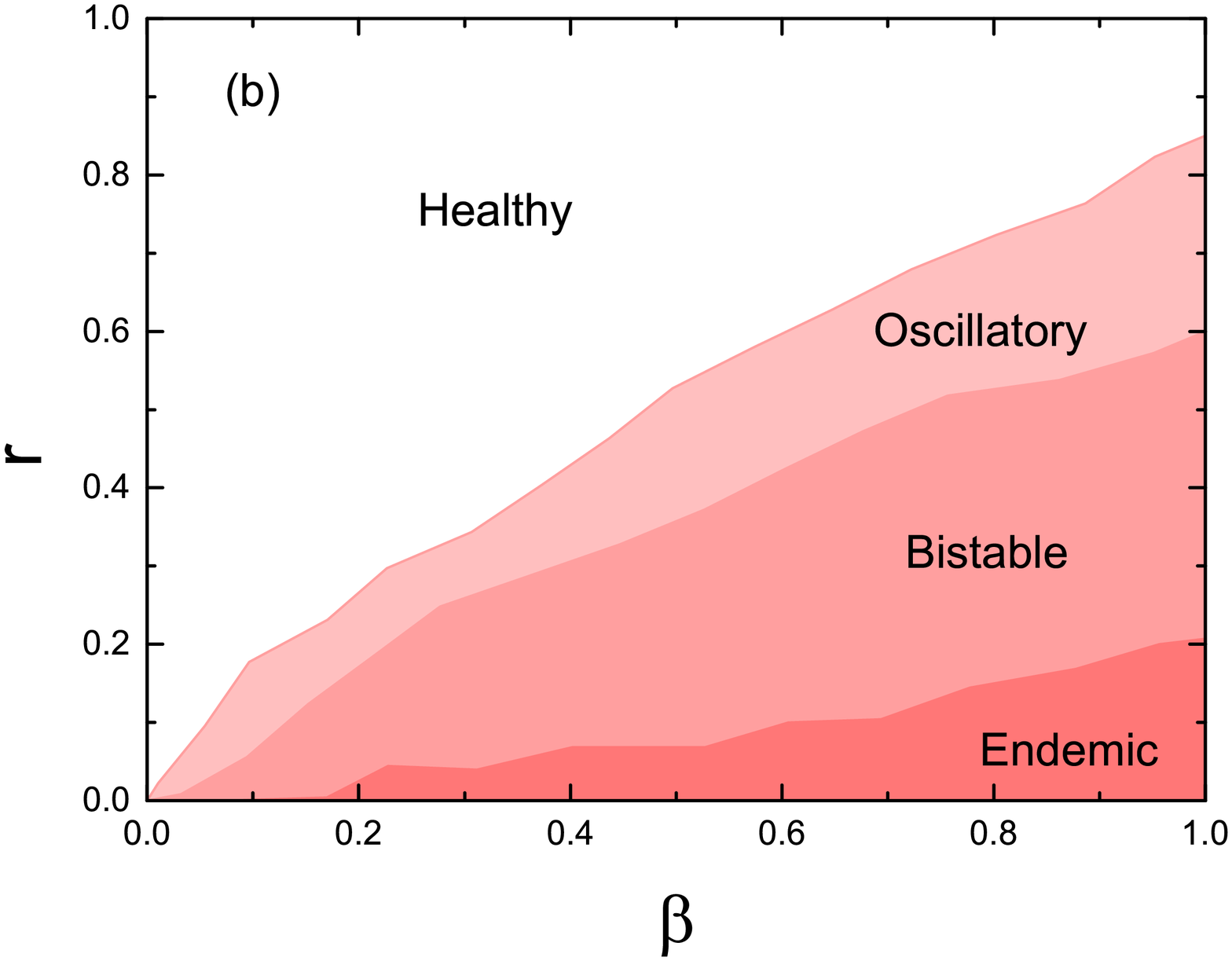}
\includegraphics[width=5.1cm]{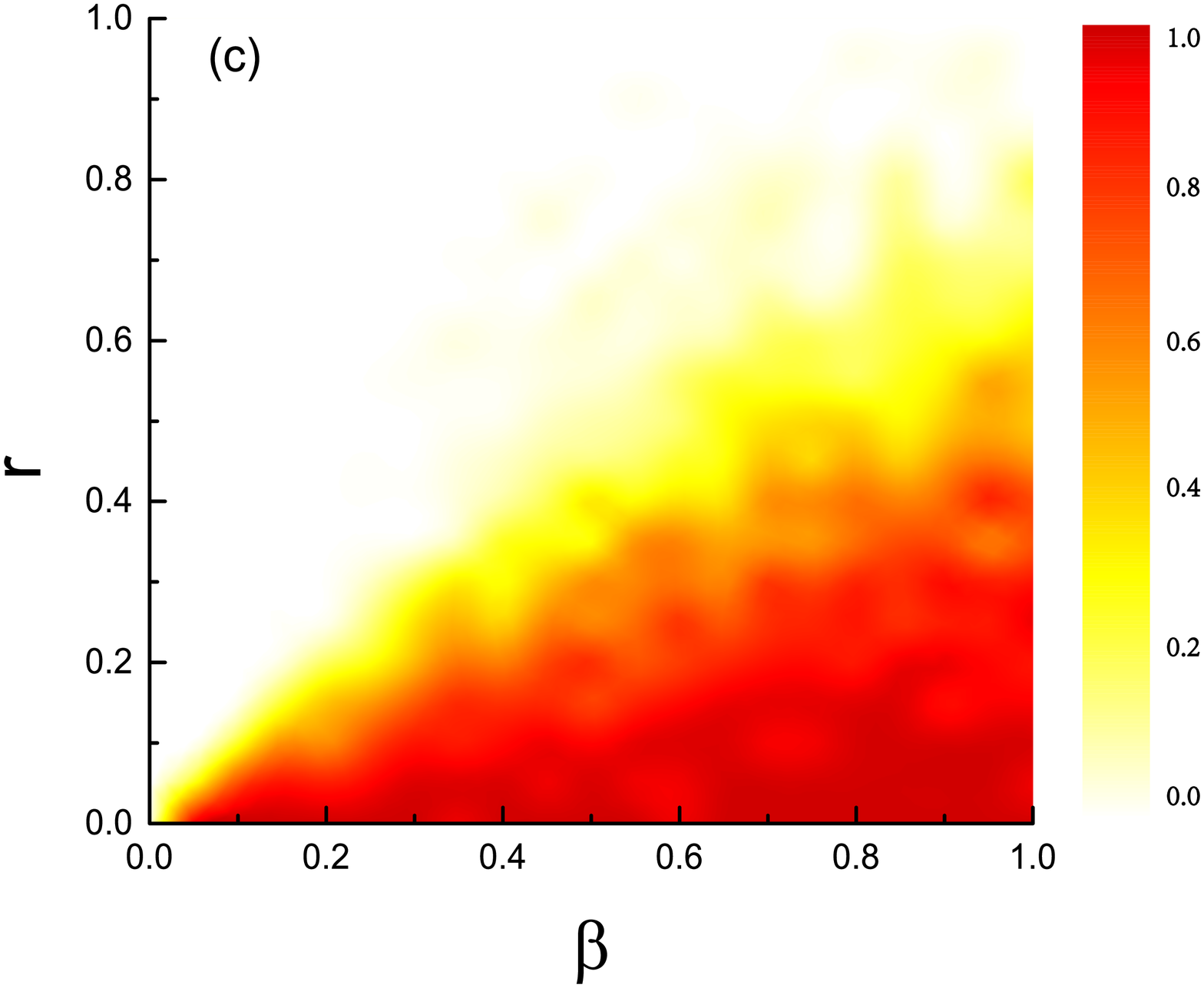}
\caption{\label{Fig9} (Color online) (a) Bifurcation diagram of the density of the infected I as a function of the infection probability $\beta$ for different values of the edge-breaking rate r based on the results of simulation of the full network (diamonds). (b) Two parameter bifurcation diagram showing the dependence on the edge-breaking rate r and the infection probability $\beta$ based on the results of simulation of the full network. (c) Full phase diagram $r-\beta$ for the simulation of the adaptive process. The parameters are the same as Fig. 8. }
\end{figure}

According to the dynamical behavior illustrated in Fig. \ref{Fig8} under different parameter sets. Bifurcation diagram of the density of the infected as a function of infected probability $\beta$ for different values of the edge-breaking rate $r$ is given by Fig. \ref{Fig9}a. Without the adaptive edge-breaking mechanism ($r=0$), the disease can spread out only if $\beta>0$ for the $SI$ process. When $r>0$, the dynamical behaviors become more complicated, where the discontinuous phase transitions, bistable, oscillatory are observed. A fast edge-breaking (large $r$) leads to a broad healthy and bistable state range (shows by the range in the arrows) in Fig. \ref{Fig9}a. In Fig. \ref{Fig9}b, we give a full $r-\beta$ bifurcation diagram according to our simulation results, and we can clearly identify the areas of healthy, oscillatory, bistability and endemic state in this model. At last, we present the dependence of the average value of infected density over 10,000 independent realizations on $r$ and $\beta$ in Fig. \ref{Fig9}c, where the changing of the density is consistent with the area classification in Fig. \ref{Fig9}b.

\section{Conclusion}
In order to understand the interplay between \emph{the dynamics on the network} (the spread of epidemic spreading and disease information) and \emph{the dynamics of the network} (the time varying of network links), we present two types of spreading dynamics with SI and SIS process respectively on an information-driven adaptive network, where the individuals who have known the disease information would probably cut off their links with others. Firstly, we illustrate the mutual feedback of epidemic spreading and information diffusion without considering the edge-breaking process ($r_S=r_I=0$), where the high epidemic prevalence preserves high disease information level, which in turn slows down the epidemic spreading. In this case, the numerical analysis based on the pairwise approach is consistent with the simulation result very well. Secondly, the results are very different when the information-driven edge-breaking process is considered ($r_S,r_I>0$). The epidemic cannot spread out if the spreading probability is smaller than the threshold (shown in Fig. \ref{Fig5}). In addition, the disease spreading and information diffusion pattern on the lattice give a visual representation that the disease might be trapped into an isolated field with information-driven adaptive process. Therefore, the information-driven adaptive process can inhibit the epidemic spreading significantly that it can not only slow down the epidemic spreading speed, but also reduce the epidemic prevalence. Finally, we give the local bifurcation analysis on four types of dynamical phenomena, including healthy, oscillatory, bistable and endemic, indicating that the state changes from healthy to oscillatory, bistable, endemic state as $\beta$ increases. %And rich dynamical phenomenon emerges with the interplay between the spreading process and the topology evaluation.

In summary, we study the dependence of the epidemic spreading on the disease information diffusion and the information-driven adaptive process, with considering the simplest spreading model (SI) and adaptive process (edge-breaking). Recent researches show the different features between the epidemic and the information diffusion \cite{Lu-Chen-Zhou-2011-NJP,Zhang-Zhang-Han-Liu-2014-PO}, and this difference dynamics would also impact the interplay between epidemic spreading and disease information diffusion significantly. Another area for future extension is to adopt other adaptation rules rather than the simple edge-breaking strategy, such as the temporarily deactivating, where the broken links would be active again after a fixed time \cite{Valdez-Macri-Braunstein-PRE-2012} or if the corresponding infected node becomes recovered \cite{Tunc-Shkarayev-Shaw-JSP-2013}.

\section*{Acknowledgments}

This work was partially supported by Natural Science Foundation of China (Grant Nos. 11305043 and 11301490), Zhejiang Provincial Natural Science Foundation of China (Grant No. LY14A050001), Zhejiang Qianjiang Talents Project (QJC1302001), and the EU FP7 Grant 611272 (project GROWTHCOM).

\section*{Appendix}
Denote $[\chi]$ as the expected values of individuals of different type described in  \textbf{Sec. Numerical mathematical analysis}, the epidemic spreading is depicted by the parameters $\beta$, $\sigma_I\beta$, $\sigma_S\beta$ and $\sigma_{SI}\beta$, while the diffusion of disease information is controlled by the parameters: $\alpha, \lambda, \omega$. All these parameters have been explained in \textbf{Sec. Model description}. According to the model described above, the the differential equations of the mean-field approach and pairwise approach are given as follows.

\noindent{\bf{Mean-field approach:}}

\begin{equation}
\label{Eq:Meanfield:All}
\left\{
\begin{split}
\frac{d[S_{-}]}{dt}=&-\langle k\rangle\beta [I_{-}][S_{-}]-\langle k\rangle\sigma_{I}\beta [I_{+}][S_{-}]-\alpha([S_{+}]+[I_{+}])[S_{-}]+\lambda [S_{+}]\\[5pt]
\frac{d[S_+]}{dt}=&-\langle k\rangle\sigma_S\beta [I_-][S_+]-\langle k\rangle\sigma_S\sigma_I\beta [I_+][S_+]+\alpha([S_+]+[I_+])[S_-]-\lambda [S_+]\\[5pt]
\frac{d[I_-]}{dt}=&\langle k\rangle\beta [I_-][S_-]+\langle k\rangle\sigma_I\beta [I_+][S_-]-\alpha([S_+]+[I_+])[I_-]-\omega [I_-]+\lambda [I_+]\\[5pt]
\frac{d[I_+]}{dt}=&\langle k\rangle\sigma_S\beta [I_-][S_+]+\langle k\rangle\sigma_S\sigma_I\beta [I_+][S_+]+\alpha([S_+]+[I_+])[I_-]+\omega [I_-]-\lambda [I_+]
\end{split}
\right.
\end{equation}

\noindent{\bf{Pairwise approach:}}
\begin{equation}
\label{Eq:Pairwise:All}
 \left\{
     \begin{split}
\frac{d[S_{-}]}{dt}=&-\beta[S_{-}I_{-}]-\sigma_{I}\beta[S_{-}I_{+}]-\alpha([S_{-}S_{+}]+[S_{-}I_{+}])+\lambda[S_{+}]\\[3pt]
\frac{d[S_{+}]}{dt}=&-\sigma_{S}\beta[S_{+}I_{-}]-\sigma_{S}\sigma_{I}\beta[S_{+}I_{+}]+\alpha([S_{-}S_{+}]+[S_{-}I_{+}])-\lambda[S_{+}]\\[3pt]
\frac{d[I_{-}]}{dt}=&\beta[S_{-}I_{-}]+\sigma_{I}\beta[S_{-}I_{+}]-\alpha([S_{+}I_{-}]+[I_{-}I_{+}])-\omega[I_{-}]+\lambda[I_{+}]\\[3pt]
\frac{d[I_{+}]}{dt}=&\sigma_{S}\beta[S_{+}I_{-}]+\sigma_{S}\sigma_{I}\beta[S_{+}I_{+}]+\alpha([S_{+}I_{-}]+[I_{-}I_{+}])+\omega[I_{-}]-\lambda[I_{+}]\\[3pt]
\frac{d[S_{-}I_{-}]}{dt}&=-\beta[S_{-}I_{-}]+\lambda([S_{+}I_{-}]+[S_{-}I_{+}])-\omega[S_{-}I_{-}]+\beta\frac{[S_{-}I_{-}]([S_{-}S_{-}]-[S_{-}I_{-}])}{[S_{-}]}+\sigma_{I}\beta\frac{[S_{-}I_{+}]([S_{-}S_{-}]-[S_{-}I_{-}])}{[S_{-}]}\\
                        &-\alpha\frac{[S_{-}I_{-}]([S_{-}I_{+}]+[S_{-}S_{+}])}{[S_{-}]}-\alpha\frac{[S_{-}I_{-}]([I_{-}I_{+}]+[S_{+}I_{-}])}{[I_{-}]}\\[3pt]
\frac{d[S_{-}I_{+}]}{dt}&=-\sigma_{I}\beta[S_{-}I_{+}]+\omega[S_{-}I_{-}]+\lambda[S_{+}I_{+}]-\alpha[S_{-}I_{+}]-\lambda[S_{-}I_{+}]-\beta\frac{[S_{-}I_{-}][S_{-}I_{+}]}{[S_{-}]}-\sigma_{I}\beta\frac{[S_{-}I_{+}]^{2}}{[S_{-}]}+\sigma_{S}\beta\frac{[S_{+}I_{-}][S_{-}S_{+}]}{[S_{+}]}\\
                        &+\sigma_{S}\sigma_{I}\beta\frac{[S_{+}I_{+}][S_{-}S_{+}]}{[S_{+}]}+\alpha\frac{[S_{-}I_{-}]([I_{-}I_{+}]+[S_{+}I_{-}])}{[I_{-}]}-\alpha\frac{[S_{-}I_{+}]([S_{-}I_{+}]+[S_{-}S_{+}])}{[S_{-}]}-r_{I}[S_{-}I_{+}]\\[3pt]
\frac{d[S_{+}I_{-}]}{dt}&=-\sigma_{S}\beta[S_{+}I_{-}]+\lambda[S_{+}I_{+}]-\lambda[S_{+}I_{-}]-\alpha[S_{+}I_{-}]-\omega[S_{+}I_{-}]-\sigma_{S}\beta\frac{[S_{+}I_{-}]^{2}}{[S_{+}]}-\sigma_{S}\sigma_{I}\beta\frac{[S_{+}I_{+}][S_{+}I_{-}]}{[S_{+}]}\\
                        &+\beta\frac{[S_{-}I_{-}][S_{-}S_{+}]}{[S_{-}]}+\sigma_{I}\beta\frac{[S_{-}I_{+}][S_{-}S_{+}]}{[S_{-}]}+\alpha\frac{[S_{-}I_{-}]([S_{-}S_{+}]+[S_{-}I_{+}])}{[S_{-}]}-\alpha\frac{[S_{+}I_{-}]([I_{-}I_{+}]+[S_{+}I_{-}])}{[I_{-}]}-r_{S}[S_{+}I_{-}]\\[3pt]
\frac{d[S_{+}I_{+}]}{dt}&=-\sigma_{S}\sigma_{I}\beta[S_{+}I_{+}]+\alpha[S_{-}I_{+}]+\alpha[S_{+}I_{-}]+\omega[S_{+}I_{-}]-2\lambda[S_{+}I_{+}]+\sigma_{S}\beta\frac{[S_{+}I_{-}]([S_{+}S_{+}]-[S_{+}I_{+}])}{[S_{+}]}\\
                        &+\sigma_{S}\sigma_{I}\beta\frac{[S_{+}I_{+}]([S_{+}S_{+}]-[S_{+}I_{+}])}{[S_{+}]}+\alpha\frac{[S_{-}I_{+}]([S_{-}I_{+}]+[S_{-}S_{+}])}{[S_{-}]}+\alpha\frac{[S_{+}I_{-}]([S_{+}I_{-}]+[I_{-}I_{+}])}{[I_{-}]}\\
                        &-[1-(1-r_{S})(1-r_{I})][S_{+}I_{+}]\\[3pt]
\frac{d[I_{-}I_{-}]}{dt}&=2\beta[S_{-}I_{-}]+2\lambda[I_{-}I_{+}]-2\omega[I_{-}I_{-}]+2\beta\frac{[S_{-}I_{-}]^{2}}{[S-]}+2\sigma_{I}\beta\frac{[S_{-}I_{+}][S_{-}I_{-}]}{[S_{-}]}-2\alpha\frac{[I_{-}I_{-}]([S_{+}I_{-}]+[I_{-}I_{+}])}{[I_{-}]}\\[3pt]
\frac{d[I_{-}I_{+}]}{dt}&=\sigma_{I}\beta[S_{-}I_{+}]+\sigma_{S}\beta[S_{+}I_{-}]+\omega([I_{-}I_{-}]-[I_{-}I_{+}])+\lambda([I_{+}I_{+}]-[I_{-}I_{+}])-\alpha[I_{-}I_{+}]\\
                        &+\beta\frac{[S_{-}I_{-}][S_{-}I_{+}]}{[S_{-}]}+\sigma_{I}\beta\frac{[S_{-}I_{+}]^{2}}{[S_{-}]}+\sigma_{S}\beta\frac{[S_{+}I_{-}]^{2}}{[S_{+}]}+\sigma_{S}\sigma_{I}\beta\frac{[S_{+}I_{+}][S_{+}I_{-}]}{[S_{+}]}+\alpha\frac{[I_{-}I_{-}]([S_{+}I_{-}]+[I_{-}I_{+}])}{[I_{-}]}\\
                        &-\alpha\frac{[I_{-}I_{+}]([S_{+}I_{-}]+[I_{-}I_{+}])}{[I_{-}]}\\[3pt]
\frac{d[I_{+}I_{+}]}{dt}&=2\sigma_{S}\sigma_{I}\beta[S_{+}I_{+}]+2\alpha[I_{-}I_{+}]+2\omega[I_{-}I_{+}]-2\lambda[I_{+}I_{+}]+2\sigma_{S}\beta\frac{[S_{+}I_{-}][S_{+}I_{+}]}{[S_{+}]}+2\sigma_{S}\sigma_{I}\beta\frac{[S_{+}I_{+}]^{2}}{[S_{+}]}\\
                        &+2\alpha\frac{[I_{-}I_{+}]([S_{+}I_{-}]+[I_{-}I_{+}])}{[I_{-}]}\\[3pt]
\frac{d[S_{-}S_{-}]}{dt}&=2\lambda[S_{-}S_{+}]-2\beta\frac{[S_{-}I_{-}][S_{-}S_{-}]}{[S_{-}]}-2\sigma_{I}\beta\frac{[S_{-}I_{+}][S_{-}S_{-}]}{[S_{-}]}-2\alpha\frac{[S_{-}S_{-}]([S_{-}S_{+}]+[S_{-}I_{+}])}{[S_{-}]}\\[3pt]
\frac{d[S_{-}S_{+}]}{dt}&=\lambda[S_{+}S_{+}]-\lambda[S_{-}S_{+}]-\alpha[S_{-}S_{+}]-\sigma_{S}\beta\frac{[S_{+}I_{-}][S_{-}S_{+}]}{[S_{+}]}-\sigma_{S}\sigma_{I}\beta\frac{[S_{+}I_{+}][S_{-}S_{+}]}{[S_{+}]}-\beta\frac{[S_{-}I_{-}][S_{-}S_{+}]}{[S_{-}]}\\
                        &-\sigma_{I}\beta\frac{[S_{-}I_{+}][S_{-}S_{+}]}{[S_{-}]}+\alpha\frac{[S_{-}S_{-}]([S_{-}S_{+}]+[S_{-}I_{+}])}{[S_{-}]}-\alpha\frac{[S_{-}S_{+}]([S_{-}S_{+}]+[S_{-}I_{+}])}{[S_{-}]}\\[3pt]
\frac{d[S_{+}S_{+}]}{dt}&=2\alpha[S_{-}S_{+}]-2\lambda[S_{+}S_{+}]-2\sigma_{S}\beta\frac{[S_{+}I_{-}][S_{+}S_{+}]}{[S_{+}]}-2\sigma_{S}\sigma_{I}\beta\frac{[S_{+}I_{+}][S_{+}S_{+}]}{[S_{+}]}+2\alpha\frac{[S_{-}S_{+}]([S_{-}S_{+}]+[S_{-}I_{+}])}{[S_{-}]}
   \end{split}
   \right.
\end{equation}

\section*{References}
\bibliographystyle{elsarticle-num}
\bibliography{Bibliography}

\newpage

\end{document}